\documentclass{aa}
\usepackage{txfonts}
\usepackage{graphicx}
\usepackage{natbib}

\begin{document}

\title{Highly Ionized Sodium X-ray Line Emission from the Solar Corona and the Abundance of Sodium}

\author{K. J. H. Phillips\inst{1} \and K. M. Aggarwal\inst{2} \and E. Landi\inst{3} \and F. P. Keenan\inst{2}}

\institute{ UCL--Mullard Space Science Laboratory, Holmbury St Mary, Dorking, Surrey RH5 6NT, United Kingdom\\
\email{kjhp@mssl.ucl.ac.uk}
\and Astrophysics Research Centre, School of Mathematics and  Physics, Queen's University Belfast, Belfast BT7 1NN, N. Ireland, United Kingdom\\ \email{K.Aggarwal@qub.ac.uk, F.Keenan@qub.ac.uk}
\and Naval Research Laboratory, Washington, D.C. 20375-5320, U.S.A.\\ \email{enrico.landi@nrl.navy.mil} }

\offprints{K. J. H. Phillips \email{kjhp@mssl.ucl.ac.uk}}

\date{Received   /Accepted     }

\abstract{The \ion{Na}{x} X-ray lines between 10.9 and 11.2~\AA\ have attracted little attention but are of interest since they enable an estimate of the coronal abundance of Na to be made. This is of great interest in the continuing debate on the nature of the FIP (first ionization potential) effect. } {Observations of the \ion{Na}{x} lines with the Solar Maximum Mission Flat Crystal Spectrometer and a rocket-borne X-ray spectrometer are used to measure the Na/Ne abundance ratio, i.e. the ratio of an element with very low FIP to one with high FIP.} {New atomic data are used to generate synthetic spectra which are compared with the observations, with temperature and the Na/Ne abundance ratio as free parameters.} {Temperature estimates from the observations indicate that the line emission is principally from non-flaring active regions, and that the Na/Ne abundance ratio is $0.07 \pm 50$\%.} {The Na/Ne abundance ratio is close to a coronal value for which the abundances of low-FIP elements (FIP $< 10$~eV) are enhanced by a factor of 3 to 4 over those found in the photosphere. For low-temperature ($T_e \leqslant 1.5$~MK) spectra, the presence of \ion{Fe}{xvii} lines requires that either a higher-temperature component is present or a revision of ionization or recombination rates is needed. }


\maketitle

\section{Introduction}

Sodium has few prominent or unblended emission lines in the ultraviolet and X-ray spectra emitted by solar coronal plasmas. Hence determinations of its abundance relative to other elements rely on measurements of emission line fluxes that are either weak or in crowded spectral regions. Yet its abundance is important for discussions of the so-called FIP effect, whereby coronal element abundances differ from corresponding photospheric values depending on whether their first ionization potential (FIP) is greater or less than about 10~eV \citep{mey85,fel92a}. This is because the FIP of Na is only 5.14~eV, less than any other element common in the Sun apart from potassium, which has an FIP of 4.34~eV. Determinations of the coronal abundance of either Na or K can therefore probe the possibility that elements with small FIP are enhanced by amounts that depend on the magnitude of the FIP, as is suggested by the electric current model of \citet{hen97} to explain the FIP effect. Observational evidence for this has been provided by \citet{fel93}, and to some extent by observations of He-like K X-ray lines seen in numerous flares and non-flare periods between 2002 and 2005 with the RESIK spectrometer \citep{syl10}. Coronal sodium emission lines include the Li-like Na (\ion{Na}{ix}) resonance lines $1s^2 2s\,\,^2S_{1/2}-1s^2\,2p\,\,^2P_{1/2}$, $^2P_{3/2}$ in the extreme ultraviolet, at 681.7~\AA\ and 694.3~\AA, and the He-like Na (\ion{Na}{x}) line $1s2s\,\,^3S_1 - 1s2p\,\,^3P_2$ at 1111.76~\AA\ seen in {\it SOHO} SUMER spectra \citep{cur00, cur01}, and resonance lines of H-like (\ion{Na}{xi}) and He-like Na (\ion{Na}{x}) in the soft X-ray range. X-ray observations of the \ion{Na}{x} and \ion{Na}{xi} X-ray emission lines should in principle be easier than for the equivalent potassium lines as the photospheric abundance of Na is higher by a factor of 16 than K \citep{asp09}, but nevertheless observations remain scarce. The \ion{Na}{xi} Ly-$\alpha$ and Ly-$\beta$ lines are at 10.02~\AA\ and 8.46~\AA\ respectively, and the \ion{Na}{x} $1s^2 - 1s2l$ ($l=s$, $p$) lines are near 11~\AA. The Ly-$\beta$ line was observed by \citet{wal74} with crystal spectrometers on the {\it OV1-17} spacecraft, and the \ion{Na}{x} lines with rocket-borne crystal spectrometers by \citet{par75}. Only a single scan of the Flat Crystal Spectrometer (FCS) on {\it Solar Maximum Mission} ({\it SMM}) in its 9-year lifetime was made of the \ion{Na}{x} lines during a flare in 1980 \citep{phi82}. Previous Na abundance analyses of the {\it OV1-17} data gave ${\rm log}\,A({\rm Na}) = 6.26$ \citep{wal74} and of the Parkinson rocket data ${\rm log}\,A({\rm Na}) = 6.73$ \citep{par75} (on a scale ${\rm log}\,A({\rm H}) = 12)$. The \ion{Na}{x} and \ion{Na}{xi} X-ray lines have incidentally been observed in radiation from a laboratory (Z-pinch) device by \citet{bur90}.

Only rather approximate atomic data were available at the time of the \citet{wal74} and \citet{par75} analyses. However, significant advances have since been made to evaluate collisional excitation rate coefficients and ionization fractions needed for the evaluation of the abundance of sodium from X-ray line fluxes. The {\sc chianti} database and software package in the Interactive Data Language (IDL) SolarSoftWare system, since its inception in the late 1990s \citep{der97, der09}, has also simplified analyses of X-ray and ultraviolet spectral data considerably. Excitation data for the \ion{Na}{x} lines presently included in {\sc chianti} are interpolated values from other elements, but more recently \citet{agg09} have calculated collisional excitation data specifically for He-like sodium with the close-coupling $R$-matrix code. Also, more refined ionization fractions assuming coronal ionization equilibrium have recently been calculated by \citet{bry09}.

The availability of new atomic data has inspired a fresh analysis, reported here, of the \ion{Na}{x} X-ray lines seen in the {\it SMM} FCS spectral scan and the spectra obtained by \citet{par75}. X-ray lines of \ion{Na}{x} were identified by \citet{phi82}, with more detailed analysis of nearby lines by \citet{lan05}. A slight revision of spectral line identifications is given here and a determination of the Na/Ne abundance ratio for this flare plasma, based on a remarkable coincidence of the contribution functions of the \ion{Na}{x} $w$ line and that of the \ion{Ne}{ix} $1s^2-1s4p$ ($w4$ or He-$\gamma$) line blended with the \ion{Na}{x} $w$ line. The Na/Ne abundance ratio is examined in the light of other recent abundance determinations and the nature of the FIP effect in solar coronal plasmas.

\section{Observations}

The {\it SMM} FCS observations \citep{phi82} were made during the decay stage of a {\it GOES} M1.5 class flare on 1980 August~25 with channel~2 of seven channels making up the instrument. Scanning flat crystals were mounted on a rotatable shaft, and radiation incident on them was via a grid collimator giving a 14~arcsecond (FWHM) field-of-view. The high-precision Baldwin drive--encoder units used to rotate the drive led to very accurate wavelengths for observed spectral lines. The whole FCS could also be scanned spatially across an emitting region on the Sun.  For the 1980 flare, a spatial scan was made over the flare emission near its onset to determine the location of the brightest point. The instrument was then pointed to this position, and a complete spectral scan taken. High-quality beryl (Be$_3$ Al$_2$ Si$_6$ O$_{18}$, $2d = 15.96$~\AA) was the diffracting crystal of channel~2, giving a spectral resolution of 0.0056~\AA\ at the wavelengths of the \ion{Na}{x} lines near 11~\AA. The scan took 17 minutes to accomplish, during which time there was a significant change in the emitting region. An analysis of the FCS data in this flare with {\sc chianti} \citep{lan05} took this into account, dividing the scan into seven time bins. The \ion{Na}{x} lines near 11~\AA\ lay at the end of time bin 1 and the start of time bin 2 of the scan, when the flare emission and temperature were still high. An emission measure analysis by \citet{lan05} using only Fe lines (stages \ion{Fe}{xvii} to \ion{Fe}{xxiii}) indicated a temperature of about 8~MK, which is reasonable for a M1.5 class flare a few minutes after its maximum emission. This analysis omitted emission lines of He-like ions such as \ion{Ne}{ix} and \ion{Mg}{xi} since they indicated a much lower temperature, of about 1--4~MK. The reason for this is that the emission measure appears to have a bimodal distribution, with a flare component ($\sim 8$~MK) and a lower-temperature component attributable to the non-flaring host active region which gives rise to the emission from the He-like ions. The \ion{Na}{x} emission lines appear to be in the latter category. Support for such emission measure distributions is provided by analysis of \ion{Ar}{xvii} X-ray lines observed during flares by the RESIK instrument \citep{syl08}; in that case, \ion{Ar}{xvii} line ratios were best fitted by a bimodal emission measure distribution having temperatures of 4.5~MK and 16~MK. Also, analysis of broad-band data from the {\it RHESSI} instrument in its A0 attenuator state \citep{phi06} similarly indicates that there is a non-flaring component of emission, with temperature corresponding to the active region, as well as a hotter flare component.

\begin{figure}
  \resizebox{\hsize}{!}{\includegraphics{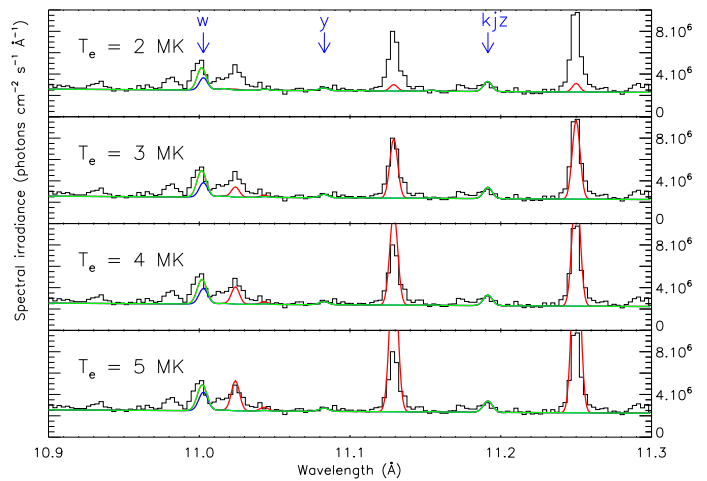}}
  \caption{X-ray spectrum in the 10.9--11.3~\AA\ range obtained with the {\it SMM} FCS during the decay of the 1980 August~25 flare (histogram) with theoretical spectra for temperatures  2~MK, 3MK, 4~MK, and 5~MK (indicated in each panel) and an assumed Ne/Na abundance ratio equal to the coronal value of 0.07 \citep{fel92a}. Line styles for the theoretical curves: continuous line = \ion{Na}{x} lines alone (including weak \ion{Na}{ix} satellites); dot-dash line = \ion{Na}{x} and \ion{Ne}{ix} lines; dotted line = with \ion{Fe}{xvii} lines. The positions of the principal \ion{Na}{x} lines $w$, $y$, and $z$ are indicated (the \ion{Na}{ix} satellites $j$ and $k$ are within 0.004~\AA\ of line $z$), and the \ion{Ne}{ix} $w4$ line is at 11.000~\AA. The other prominent lines are due to \ion{Fe}{xvii}, with observed wavelengths 11.023~\AA, 11.133~\AA, and 11.253~\AA\ (see Table~\ref{principal_lines}). (A colour version is available in the on-line journal. On-line journal version key: black histogram = observed spectrum; coloured curves are theoretical spectra, with code blue = \ion{Na}{x} lines alone; green = \ion{Na}{x} and \ion{Ne}{ix} lines; red = with \ion{Fe}{xvii} lines.) }
  \label{FCS_synth_sp}
\end{figure}

The FCS channel~2 spectral scan over the 10.9--11.3~\AA\ range is shown in each of the four panels of Figure~\ref{FCS_synth_sp} (histogram plot). In these plots, FCS count rates have been converted to absolute spectral irradiance units using pre-launch intensity calibration factors. The background emission is largely due to fluorescence of the crystal material with the solar continuum making a minor contribution. The theoretical wavelengths of  \ion{Na}{x} lines are indicated.

\begin{figure}
\resizebox{\hsize}{!}{\includegraphics{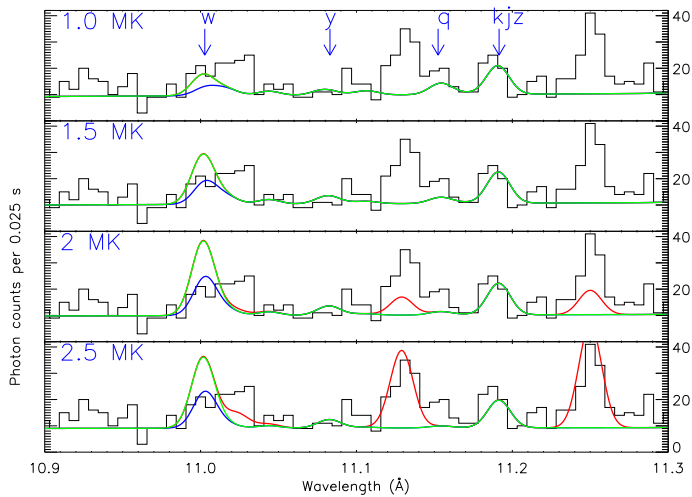}}
  \caption{X-ray spectrum (histograms) in the 10.9--11.3~\AA\ range obtained with a rocket-borne KAP  crystal spectrometer of a non-flaring active region in 1971 \citep{par75}. The spectrum is compared with synthetic spectra having temperatures 1~MK, 1.5~MK, 2~MK, and 2.5~MK (indicated in each panel) and an assumed Ne/Na abundance ratio equal to a coronal value of 0.07. \ion{Na}{ix} satellites are included in the synthetic spectra. Line styles are as for Figure~\ref{FCS_synth_sp}. (A colour version is available in the on-line journal. On-line journal version key: black histogram = observed spectrum; coloured curves are theoretical spectra, with code blue = \ion{Na}{x} lines alone; green = \ion{Na}{x} and \ion{Ne}{ix} lines; red = with \ion{Fe}{xvii} lines.)}
  \label{Parkinson_synth_sp}
\end{figure}

The rocket-borne crystal spectrometers described by \citet{par75} were launched on a stabilised Skylark rocket on 1971 November~30. The instruments viewed a non-flaring active region on the Sun through a multi-grid collimator with field of view equal to 3~arcminutes (FWHM). Two of the spectrometers scanned through the region of the \ion{Na}{x} lines near 11~\AA\ with KAP ($2d = 26.64$~\AA) and gypsum ($2d = 15.19$~\AA) crystal. As the data are no longer in digital form, the spectra in the region of the \ion{Na}{x} lines (10.9--11.4~\AA) from Figure~3 of \citet{par75} were hand-digitised for the purposes of this analysis. The photon count rate is higher in the KAP crystal scan and so the spectrum has high statistical quality. This scan is shown in Figure~\ref{Parkinson_synth_sp}. The temperature is likely to be lower than the FCS scan in Figure~\ref{FCS_synth_sp} and the line feature at 11.152~\AA, if real, could be the \ion{Na}{ix} satellites $q$ and $r$; the theoretical position of line $q$ is indicated.

\section{Atomic data and line identifications}

\subsection{The spectrum in the 10.9--11.3~\AA\ region}

\ion{Na}{x} lines in the X-ray region include a group of four lines with transitions of the type $1s^2 - 1s2l$ ($l=s$, $p$) and higher members of the $1s^2 - 1snl$ series. Using the notation of \citet{gab72}, the $n=2$ lines are, in increasing order of wavelength, identified with letters $w$ (transition $1s^2\,^1S_0 - 1s2p\,^1P_1$, wavelength 11.003~\AA), $x$ ($1s^2\,^1S_0 - 1s2p\,^3P_1$, 11.080~\AA), $y$ ($1s^2\,^1S_0 - 1s2p\,^3P_2$, 11.083~\AA), and $z$ ($1s^2\,^1S_0 - 1s2s\,^3S_1$, 11.192~\AA). The wavelengths, quoted from {\sc chianti}, are from \citet{ral08}.  Dielectronic satellites with transitions $1s^2 n'l' - 1s2p n'l'$ occur in great profusion in the neighbourhood of the $w$--$z$ lines, but with only a few that are significant at lower temperatures, particularly satellites $j$ and $k$ (transitions $1s^2 2p - 1s2p^2$) which blend with line $z$ and satellites $q$ and $r$ ($1s^2 2s - 1s 2s 2p$). Numerous satellites with very high $n$ values occur near line $w$ and converge on it; individually they are not important but their cumulative contribution is up to about 10\% of line $w$.

The \ion{Ne}{ix} $1s^2\,^1S_0 - 1s4p\,^1P_1$ ($w4$) line occurs very near the \ion{Na}{x} $w$ line, but how near was unclear from early atomic structure calculations. Those of \citet{vai78} gave the wavelength as 11.025~\AA, longward of the \ion{Na}{x} $w$ line by an amount easily resolvable with the FCS and Parkinson instruments. Using the wavelengths of \citet{wei69}, \citet{par75} was led to identify a line feature at 11.027~\AA\ as the \ion{Ne}{ix} $w4$ line. However, the more recent MCDF calculations of \citet{che06} seem to have established that this line is at 11.000~\AA, just 0.003~\AA\ away from the \ion{Na}{x} $w$ line. In light of this, the feature at 11.027~\AA\ is now identifiable with an \ion{Fe}{xvii} line with transition $2s^2\,2p^6\,^1S_0 - 2s\,2p^6\,4p\,^1P_1$ (wavelength 11.023~\AA\ according to \citet{lan05}). More intense \ion{Fe}{xvii} lines occur at 11.133~\AA\ and 11.253~\AA. Table~\ref{principal_lines} lists identifications and wavelengths of the principal lines in this region.

\begin{table*}
\caption{Principal lines in the 10.9--11.3~\AA\ region. } \label{principal_lines} \centering
\begin{tabular}{l l l l l}
\hline\hline
Wavelength (\AA)$^{(a)}$ & Ion &  Line label & Transition  \\

11.000 & Ne IX & $w4$ & $1s^2\,  ^1S_0 - 1s 4p\,\, ^1P_1$  \\

11.003 & Na X & $w$ & $1s^2\,  ^1S_0 - 1s 2p\,\, ^1P_1$  \\

11.023 & Fe XVII & & $2s^2\,2p^6\,^1S_0 - 2s\,2p^6\,4p\,\,^1P_1$ \\

11.080 & Na X & $x$ & $1s^2\,  ^1S_0 - 1s 2p\,\, ^3P_2$  \\

11.083 & Na X & $y$ & $1s^2 \, ^1S_0 - 1s 2p\,\, ^3P_1$  \\

11.133 & Fe XVII & & $2s^2\,2p^6\,\,^1S_0 - 2s^2\,2p^5\,5d\,\,^1P_1$ \\

11.192 & Na X & $z$ & $1s^2 \, ^1S_0 - 1s 2s\,\, ^3S_1$  \\

11.253 & Fe XVII & & $2s^2\,2p^6\,\,^1S_0 - 2s^2\,2p^5\,5d\,\,^3D_1$ \\

\hline

\end{tabular}

Note: $(a)$ Wavelengths from \citet{lan05}.

\end{table*}

\ion{Ne}{viii} satellite lines with transitions $1s^2\,nl - 1s\,nl\,4p$ are expected to form weak features to the long-wavelength side of the \ion{Ne}{ix} $w4$ line, but are outside the range of interest here: the most prominent are those in the $1s^2\,2p - 1s\,2p\,4p$ array which are located at $\sim 11.45$~\AA.
\subsection{Contribution functions}

\begin{figure}
\resizebox{\hsize}{!}{\includegraphics{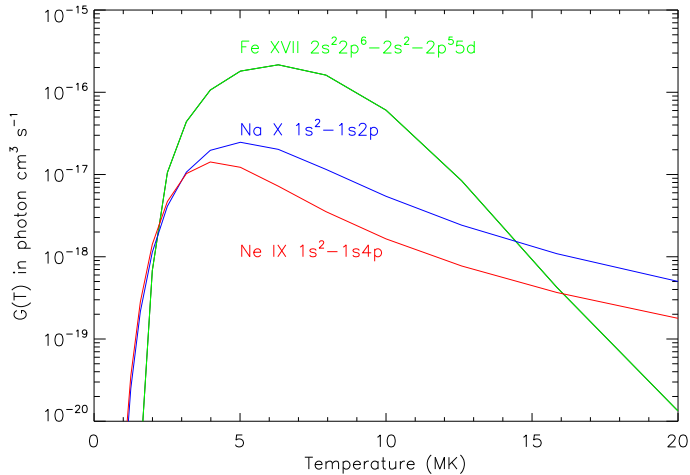}}
  \caption{Contribution functions $G(T_e)$ for lines in the 10.9--11.2~\AA\ range: \ion{Na}{x} $w$ $1s^2\,^1S_0 - 1s2p\,^1P_1$ 11.003~\AA; \ion{Ne}{ix} $w4$ $1s^2\,^1S_0 - 1s4p\,^1P_1$ 11.000~\AA; \ion{Fe}{xvii} $2s^2 2p^6\,^1S_0 - 2s^2 2p^5 5d\,^1P_1$ 11.129~\AA. The contribution functions of the \ion{Na}{x} and \ion{Ne}{ix} lines are almost identical for $T_e < 3$~MK. (A colour version is available in the on-line journal. On-line journal version key:  blue curve = \ion{Na}{x} $w$ line; red = \ion{Ne}{ix} $w4$ line; green = \ion{Fe}{xvii} 11.129~\AA\ line.) }
  \label{GofT_functions}
\end{figure}

The contribution function $G(T_e)$, defining the temperature range of significant emission of a line emitted by an ion (stage $+m$) of element X in a transition from an excited level $j$ to the ground level, is defined by (e.g. \citet{phi08})
\begin{equation}
G(T_e) =  \frac{N({\rm X}^{+m}_j)}{N({\rm X}^{+m})} \frac{N({\rm X}^{+m})}{N({\rm X})} \frac{N({\rm X})}{N({\rm H})} \frac{N({\rm H})}{N_e} N_e
\end{equation}

\noindent where the number densities $N$ are of the excited level of ion X$^{+m}_j$, the ion X$^{+m}$ (all levels summed), the element X (all ionization stages $m$), and hydrogen (H), and $N_e$ is the electron density. The abundance of X relative to H is $N({\rm X})/N({\rm H}) = A({\rm X})$, and for a coronal plasma $N({\rm H})/N_e$ is 0.8. These functions can be determined from {\sc chianti} with the user supplying the line wavelength and a chosen set of ionization fractions and element abundances. Figure~\ref{GofT_functions} shows the {\sc chianti} $G(T_e)$ functions for the lines of interest here, viz. \ion{Na}{x} $w$, \ion{Ne}{ix} $w4$, and the \ion{Fe}{xvii} line at 11.129~\AA. The excitation data for the \ion{Na}{x} lines used in these calculations are from interpolation of distorted-wave calculations for neighbouring ions, while those for the \ion{Ne}{ix} lines are from the $R$-matrix calculations of \citet{che06}. The recent ionization fractions of \citet{bry09} were used, as were the coronal abundances of \citet{fel92a}, in which low-FIP elements have abundances enhanced over photospheric values by a factor of $\sim 4$ but high-FIP element abundances are equal to photospheric values (see \S 3.5).

While the \ion{Fe}{xvii} $G(T_e)$ curve steeply rises at $T_e \lesssim 4$~MK and falls at $T_e \gtrsim 10$~MK, reflecting the fractional abundance of the Fe$^{+16}$ ion, both the \ion{Na}{x} and \ion{Ne}{ix} lines have contribution functions that are much more slowly decreasing with $T_e$. There is a remarkable coincidence of the \ion{Na}{x} and \ion{Ne}{ix} curves for $T_e \lesssim 4$~MK, and even at higher temperatures the \ion{Na}{x} curve is consistently only a factor 3 above the \ion{Ne}{ix} curve. This provides a means for determining the Na/Ne abundance if the two lines, which are practically unresolvable with the FCS and Parkinson spectrometers, are mostly emitted at temperatures $\lesssim 4$~MK.

\subsection{Theoretical Na X line fluxes}

The principal \ion{Na}{x} X-ray lines in the 10.9--11.3~\AA\ range are mostly excited by electron collisions, and so collisional excitation rate coefficients are of considerable importance in synthesizing the spectra. As mentioned, the {\sc chianti} data use interpolated distorted-wave calculations, but recent excitation data of \citet{agg09} using the close-coupling $R$-matrix code developed at Queen's University Belfast should be a significant improvement and were used here. The data are in the form of temperature-averaged collision strengths $\Upsilon (T_e)$ for the various transitions involved, and include auto-ionizing resonances which make a contribution to the $\Upsilon$s if near the excitation threshold. For the \ion{Na}{x} $w$ line, the excitation is mostly from the ground state $1s^2\,\,^1S_0$ to the upper level $1s2p\,\,^1P_1$, with spontaneous radiative transition from the upper level. However, for the $x$, $y$, and $z$ lines in which the upper levels are triplets ($1s2s\,\,^3S_1$ or $1s2p\,\,^3P$), other transitions are involved. Cascade transitions from higher levels $1s\,nl$ where $n>3$ are important, especially for lines $x$, $y$, and $z$, as are transitions from the upper levels of each line resulting from recombination of the H-like stage of Na (Na$^{+10}$). The new excitation data were entered into {\sc chianti} data files so that \ion{Na}{x} line fluxes could be obtained. While the differences in  the $x$, $y$, and $z$ line fluxes between the original interpolated data from {\sc chianti} and the new $R$-matrix results are rather modest ($\lesssim 7$~\%), they are up to 22~\% ($T_e=3$~MK) for the important $w$ line.

\subsection{Na IX dielectronic satellites}

Dielectronic satellite line emission is an important contributor to X-ray spectra of He-like and H-like ions, particularly those with high atomic number $Z$. The {\sc chianti} database currently has no dielectronic satellite excitation data for satellites due to Li-like Na which occur near the \ion{Na}{x} X-ray lines, having transitions $1s^2 nl - 1s2p nl$ ($nl$ being the quantum numbers of the ``spectator" electron in the transition: see \citet{gab72}). Previous calculations have used a variety of atomic codes to obtain these data; here we used an adaptation of the Cowan pseudo-relativistic Hartree-Fock (HFR) code \citep{cow81} for small personal computers designed by A. Kramida (2008, priv. comm.).

The data concerned are the satellite (sat) wavelength and an intensity factor $B_{\rm sat}$. A satellite line is excited by dielectronic recombination when an electron recombines with a He-like Na ion (Na$^{+9}$) to produce a doubly excited state which then radiatively de-excites. The flux at the Earth (distance from Sun = 1 AU) of the satellite $F_{\rm diel}$ formed in this way  is given by

\begin{equation}
F_{\rm diel} = 2.07 \times 10^{-16} \frac{N({\rm Na}^{+9}) N_e V}{4\pi ({\rm AU})^2} \frac{A_r A_a}{A_a + \Sigma A_r} \frac
{{\rm exp} \, (-E_{\rm sat} / k_B T_e)} {T_e^{3/2}}
\end{equation}

\noindent where $A_r$ and $A_a$ are the radiative and autoionization probabilities from the satellite's upper level respectively, $E_{\rm sat}$ the energy of the upper level with respect to the ground level of the Na$^{+9}$ ion, $V$ the flare emitting volume, and $k_B$ is Boltzmann's constant. The term $ N({\rm Na}^{+9}) N_e V$ can be rewritten $0.8 f({\rm Na}^{+9}) A({\rm Na}) EM$ where the volume emission measure is $EM = N_e^2 V$ and the fraction of Na$^{+9}$ ions, $N({\rm Na}^{+9})/\Sigma N({\rm Na}^{+m})$, is $f(T_e)$. The intensity factor $B_{\rm sat}$, defined by

\begin{equation}
B_{\rm sat} = \frac{A_r A_a}{A_a + \Sigma A_r}\,{\rm ,}
\end{equation}

\noindent is calculated by the Cowan HFR code together with all the radiative and autoionization probabilities and the satellite wavelengths.

Input to the code is the satellite array and value of $E_{\rm sat}$. Single-electron radial functions are calculated by the code and combined with Slater--Condon theory to give energy levels. Previous experience \citep{phi94} has shown that choosing 100\% for the scaling of the Slater parameters in the code gives good results for X-ray transitions, though there is generally a small wavelength shift $\Delta \lambda$ between the wavelength of the He-like ion line $w$ and that of satellites with very high values of $n$ for the spectator electron, which should converge on line $w$. In running the code for \ion{Na}{ix} satellites, we took 100\% scaling of the Slater parameters and added $\Delta \lambda = + 0.001$~\AA\ to the Cowan satellite wavelengths since this value of $\Delta \lambda$ is needed to achieve agreement between the wavelengths of high-$n$ satellites and the \ion{Na}{x} $w$ line.

Table~\ref{sat_lines} lists data for a selection of satellites -- wavelengths, $B_{\rm sat}$, and $E_{\rm sat}$ -- from the calculations, which included satellites with spectator electrons having $nl = 2s$, $2p$, $3s$, $3p$, $3d$, etc., extending up to $nl = 6p$. We generated data for $nl=10p$ to determine the wavelength shift $\Delta \lambda$. As was found by \citet{gab72} and others, satellites $j$ and $k$ (see Table~\ref{sat_lines}) are the most intense, i.e. have the largest values of $B_{\rm sat}$. They are both within 0.004~\AA\ of the \ion{Na}{x} $z$ line at 11.192~\AA. This is in agreement with other calculations of satellite line wavelengths for elements having similar atomic number, notably Mg (e.g. \citet{ste84}). For  Ca ($Z = 20$), satellite $k$ becomes resolved from line $z$, while for Fe ($Z=26$), both $j$ and $k$ are resolved.

\begin{figure}
\resizebox{\hsize}{!}{\includegraphics{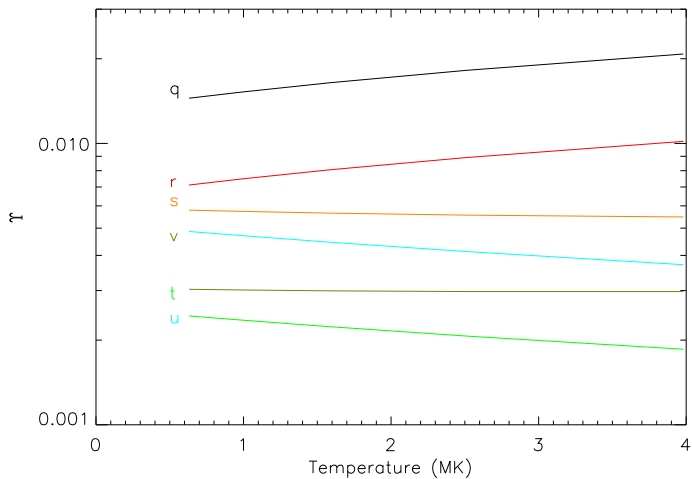}}
  \caption{Maxwellian-averaged collision strengths $\Upsilon (T_e)$ as a function of temperature for excitation from the Li-like Na (Na$^{+8}$) ion ground level $1s^2 2s\,\,^2S_{1/2}$ to the upper levels of satellites $q$, $r$, $s$, $t$, $u$, and $v$ (transitions given in Table~\ref{sat_lines}) calculated using the Flexible Atomic Code (FAC). (A colour version is available in the on-line journal, with different colours for the $\Upsilon$ curves for excitation to each upper level.)}
  \label{Upsilon_plot}
\end{figure}

Inner-shell excitation can also give rise to satellites in the $1s^2 2s - 1s 2s 2p$ array for solar flare densities (at much higher densities, inner-shell excitation gives rise to satellites in other arrays also). Again, the current version of {\sc chianti} does not have atomic data relating to these satellites. As at least satellites $q$ and $r$ (see Table~\ref{sat_lines}) in the $1s^2 2s - 1s 2s 2p$ array are expected to be important, Maxwellian-averaged collision strengths $\Upsilon(T_e)$ for the entire array were calculated using the Flexible Atomic Code of \citet{gu03}. Assuming that the de-excitation is entirely the radiative transition to the ground level $1s^2 2s\,\,^2S_{1/2}$, the satellite line fluxes are then given by

\begin{equation}
F_{\rm i-s} = \frac{N({\rm Na}^{+8}) N_e V}{4\pi ({\rm AU})^2} \times \frac{8.63 \times 10^{-6} \Upsilon (T_e)}{T_e^{1/2}} \,\,{\rm exp}\,\,(-E_{\rm exc} / k_B T_e)
\end{equation}

\noindent where $E_{\rm exc}$ is the excitation energy of the satellite line. Figure~\ref{Upsilon_plot} shows the calculated $\Upsilon$ functions for excitation from the $1s^2 2s\,\,^2S_{1/2}$ level to the levels indicated.

\begin{table*}
\caption{Selected \ion{Na}{ix} satellite lines excited by dielectronic recombination: data from the Cowan HFR code.  } \label{sat_lines} \centering
\begin{tabular}{l l l l l}
\hline\hline
 Wavelength (\AA)$^{(a)}$ &  Line label$^{(b)}$ & Transition & $B_{\rm sat}$ (s$^{-1}$)$^{(c)}$ & $E_{\rm sat}$ (Ry)  \\

11.043 & $d15$ & $1s^2 3p\, ^2P_{1/2} - 1s 2p 3p\, (^1P)\,^2D_{3/2}$ & 3.33(13) & 73.5 \\

11.044 & $d13$ & $1s^2 3p\, ^2P_{3/2} - 1s 2p 3p\, (^1P)\,^2D_{5/2}$ & 5.77(13) & 73.5 \\

11.077$^{(d)}$ & $s$ & $1s^2 2s\, ^2S_{1/2} - 1s 2p^2\, (^3S)\,^2P_{3/2}$ & 4.64(12) & 60.0 \\

11.078$^{(d)}$ & $t$ & $1s^2 2s\, ^2S_{1/2} - 1s2s2p\,(^3S)\,^2P_{1/2}$ & 2.90(12) & 60.0 \\

11.105 & $n$ &$1s^2 2p\, ^2P_{1/2} - 1s 2p^2\, ^2S_{1/2}$ & 3.11(12) & 60.6 \\

11.108 & $m$ &$1s^2 2p\, ^2P_{3/2} - 1s 2p^2\, ^2S_{1/2}$ & 8.19(12) & 60.6 \\

11.152$^{(d)}$ & $q$ & $1s^2 2s\, ^2S_{1/2} - 1s2s2p\,(^1S)\,^2P_{3/2}$ & 1.66(13) & 60.6 \\

11.155$^{(d)}$ & $r$ & $1s^2 2s\, ^2S_{1/2} - 1s2s2p\,(^1S)\,^2P_{1/2}$ & 9.61(12) & 60.6 \\

11.165 & $b$ &$1s^2 2p\, ^2P_{1/2} - 1s 2p^2\, ^2P_{3/2}$ & 3.28(11) & 60.6 \\

11.168 & $d$ &$1s^2 2p\, ^2P_{1/2} - 1s 2p^2\, ^2P_{1/2}$ & 7.12(10) & 60.6 \\

11.168 & $a$ &$1s^2 2p\, ^2P_{3/2} - 1s 2p^2\, ^2P_{3/2}$ & 2.13(12) & 60.6 \\

11.171 & $c$ &$1s^2 2p\, ^2P_{3/2} - 1s 2p^2\, ^2P_{1/2}$ & 3.25(10) & 60.6 \\

11.188 & $k$ & $1s^2 2p\, ^2P_{1/2} - 1s 2p^2\, ^2D_{3/2}$ & 2.31(13) & 60.6 \\

11.191 & $j$ &$1s^2 2p\, ^2P_{3/2} - 1s 2p^2\, ^2D_{5/2}$ & 3.70(13) & 60.6 \\

11.192 & $l$ &$1s^2 2p\, ^2P_{3/2} - 1s 2p^2\, ^2D_{3/2}$ & 1.90(12) & 60.6 \\

11.286 & $e$ & $1s^2 2p\, ^2P_{3/2} - 1s 2p^2\, ^4P_{5/2}$ & 1.74(10) & 60.6 \\

11.298$^{(d)}$ & $u$ & $1s^2 2s\, ^2S_{1/2} - 1s 2p^2\, (^3S)^4P_{3/2}$ & 3.45(9) & 60.0 \\

11.300$^{(d)}$ & $v$ & $1s^2 2s\, ^2S_{1/2} - 1s2s2p\, (^3S)^4P_{1/2}$ & 6.64(8) & 60.0 \\

\hline

\end{tabular}

Note: $(a)$ Wavelengths from Cowan code + 0.001~\AA; $(b)$ Notation of \citet{gab72,bel79}; $(c)$ Numbers in parentheses are powers of ten; $(d)$ Lines also formed by inner-shell excitation.

\end{table*}

\subsection{Synthetic spectra}

A spectral synthesis program was written with the specific intention of matching the \ion{Na}{x} and \ion{Ne}{ix} lines, together with \ion{Na}{ix} satellites, in the 10.9--11.3~\AA\ region as observed by the FCS and Parkinson spectrometer. It is assumed, as was found from the RESIK flare observations of \ion{Ar}{xvii} line ratios \citep{syl08} and analysis of broad-band spectra from {\it RHESSI} in its A0 attenuator state \citep{phi06}, that a two-component emission measure distribution describes the observed spectra, a low-temperature component (1--4~MK) appropriate for the \ion{Na}{x} and \ion{Ne}{ix} emission from the non-flaring active region and a higher-temperature component for the flare emission proper (FCS spectrum) or for a hotter part of the active region if present (Parkinson spectra): it is this component that is responsible for most of the Fe ion line emission in this region, including \ion{Fe}{xix} and \ion{Fe}{xxiii} lines seen in the FCS spectrum and which were fitted by an 8~MK component in the analysis by \citet{lan05}. The \ion{Fe}{xvii} lines are emitted by plasma with $T_e > 3$~MK (see Figure~\ref{GofT_functions}) so contributions from both the active region and flare may be expected. The synthesis program computed spectra in a temperature grid from 0.8~MK to 5~MK.

The \ion{Na}{x} lines were included in the synthesis program from the \citet{agg09} atomic data (\S 3.3) and the \ion{Na}{ix} satellites, both those excited by dielectronic recombination and those formed by inner-shell excitation, from the data discussed in \S 3.4. The \ion{Ne}{ix} $w4$ line was included using fluxes from {\sc chianti}, as well as the much weaker intercombination line $1s^2\,\,^1S_0-1s4p\,\,^3P_1$, unresolvable from the $w4$ line. Other lines were also included, most notably \ion{Fe}{xvii} lines; lines of \ion{Fe}{xix} and \ion{Fe}{xxiii} had negligible fluxes in the temperature range considered. The {\sc chianti} \ion{Fe}{xvii} line wavelengths differ somewhat from the observed FCS wavelengths; as the FCS wavelengths are expected to be very precise (uncertainties $\lesssim 2$~m\AA\ in this region), the FCS wavelengths from an original analysis \citep{phi82} were used. Fluxes of the \ion{Na}{x} $w$, $x$, $y$, and $z$ lines were obtained from the collisional excitation data of \citet{agg09} which were inserted into files that could be read by the {\sc chianti} software. The \ion{Na}{ix} satellite fluxes were calculated from equations~(2) and (4). Gaussian profiles were applied to spectral lines that were a convolution of the thermal Doppler broadening (FWHM width $\Delta \lambda_D$) and the instrumental profile, defined by the crystal rocking curve (FWHM width equal to $\Delta \lambda_{\rm rc}$), assumed to be Gaussian.  The widths are given respectively by

\begin{equation}
\Delta \lambda_D = 1.665\,\, \frac{\lambda}{c} \,\,\left (\frac{2 k_B T_{\rm ion}}{M_{\rm ion}}\right )^{1/2}
\end{equation}

\noindent where $T_{\rm ion}$ is the temperature of the emitting ion, taken to be $T_e$, and $M_{\rm ion}$ the ion mass, and

\begin{equation}
\Delta \lambda_{\rm rc} = 2d \,\,{\rm cos} \,\,\theta \,\,\Delta \theta
\end{equation}

\noindent where the crystal rocking curve is $\Delta \theta$. A pre-launch measured value of $\Delta \theta$ for the FCS observing the \ion{Na}{x} lines of 100~arcseconds was taken for the FCS spectrum. The differing ion masses for the \ion{Na}{x}, \ion{Ne}{ix}, and \ion{Fe}{xvii} lines were taken into account, though the FCS instrumental broadening dominates the convolved line profile. For the Parkinson rocket spectra, the instrument profile was empirically taken to be 3 times the FCS rocking curve width to match the observed line profiles.  As in the calculation of contribution functions (\S 3.2), the Na and Ne ionization fractions were taken from \citet{bry09}.

Element abundances were taken for an average coronal plasma using values from \citet{fel92a} (included as ``coronal abundances" in {\sc chianti}). An initial value of 0.071 is thus taken for the sodium-to-neon abundance ratio -- $A({\rm Na})/A({\rm Ne})$ -- but in our analysis it is a free parameter to be determined from line flux measurements. The Fe abundance likewise is taken to be ${\rm log}\,A({\rm Fe}) = 8.10$ (on a scale ${\rm log}\,A({\rm H}) = 12$. This is supported by measurements of broad-band {\it RHESSI} flare spectra \citet{phi06}, indicating that of 27 flares, 19 had estimated abundances  ${\rm log}\,A({\rm Fe})$ that were within 20\% of the \citet{fel92a} value. Although spatial variations in coronal abundances have been noted \citep{fel92b}, the chief variations are those during impulsive flares and above strong sunspot magnetic fields, when low-FIP elements appear to have photospheric abundances.

\section{Results}

Figure~\ref{FCS_synth_sp} shows the {\it SMM} FCS spectrum for the flare of 1980 August~25. In this figure, the observed spectrum is compared with four synthetic spectra with $T_e = 2$, 3, 4, 5~MK, all with the coronal value of $A({\rm Na})/A({\rm Ne}) = 0.071$. In each case, the synthetic spectra were adjusted so that the blend of the \ion{Na}{x} $z$ line with the \ion{Na}{ix} $j$ and $k$ satellites (feature at 11.192~\AA) is fitted, leaving the blend of the \ion{Na}{ix} $w$ line and the \ion{Ne}{ix} $w4$ line (feature at 11.003~\AA) free as well as the barely significant \ion{Na}{x} $y$ line at 11.083~\AA. At the temperatures shown, the \ion{Na}{ix} satellites $j$ and $k$ are very weak, the ratio of the sum of these two satellites to the \ion{Na}{x} $z$ line varying from 0.24 to 0.045 over the range $T_e = 2$~MK to 5~MK. The satellites $q$ and $r$ are also weak, the feature at 11.15~\AA\ formed by them being unobserved by the FCS. In addition the ratio of blended \ion{Na}{ix} $w$ to the \ion{Ne}{ix} $w4$ lines is only weakly $T_e$-dependent (Figure~\ref{GofT_functions}). In fact, the only indication of temperature in the theoretical spectra in Figure~\ref{FCS_synth_sp} is the presence of nearby \ion{Fe}{xvii} lines at 11.023~\AA, 11.129~\AA, and 11.250~\AA. The best-fit temperature to the FCS spectrum is clearly $T_e = 3$~MK. However, the emission measure analysis of \citet{lan05} shows that much of the emission of these and other more highly ionized Fe lines for this particular stage of the August~25 flare occurs at a higher temperature, around $T_e \sim 8$~MK, with He-like ion line emission such as \ion{Na}{x} and \ion{Ne}{ix} occurring at lower temperatures indicative of the host active region. The agreement of the FCS and theoretical spectra at $T_e = 3$~MK for the \ion{Fe}{xvii} lines in Figure~\ref{FCS_synth_sp} may therefore be regarded as coincidental.

\begin{figure}
\resizebox{\hsize}{!}{\includegraphics{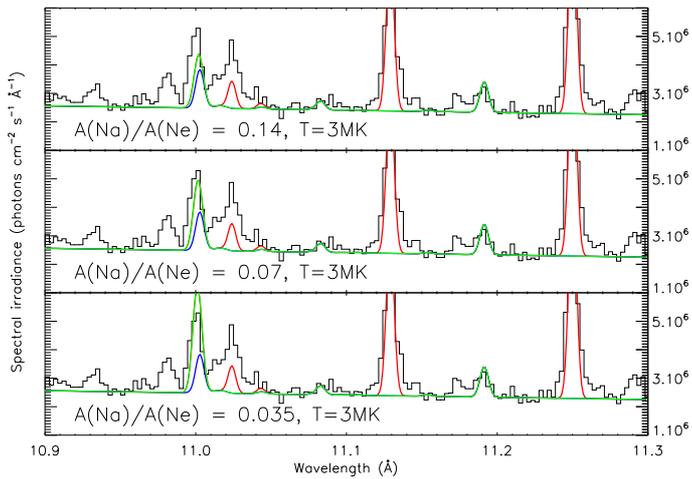}}
  \caption{The FCS spectrum in the 10.9--11.3~\AA\ range during the flare of 1980 August~25 compared with theoretical spectra for $T_e = 3$~MK and three values of the abundance ratio $A({\rm Na})/A({\rm Ne})$ (indicated in each plot). Line styles for the theoretical curves as for Figure~\ref{FCS_synth_sp}. Two of the \ion{Fe}{xvii} lines are off-scale to show better the agreement of the theoretical curves with the observed \ion{Na}{x} and \ion{Ne}{ix} line emission. (A colour version is available in the on-line journal. On-line journal version key is the same as for Figure~\ref{FCS_synth_sp}.) }
  \label{FCS_abund}
\end{figure}

As values of temperature over the range 2--5~MK only slightly affect the \ion{Na}{x} and \ion{Ne}{ix} line emission, the effect of the abundance ratio $A({\rm Na})/A({\rm Ne})$ may be examined; this is shown in Figure~\ref{FCS_abund}. The FCS spectrum is compared with theoretical spectra, calculated at $T_e = 3$~MK, for three values of $A({\rm Na})/A({\rm Ne})$: 0.14, 0.07, 0.035, i.e. the coronal value of \citet{fel92a} multiplied by 2, 1, and 0.5 respectively. Clearly the best match is for $A({\rm Na})/A({\rm Ne}) = 0.07$. Ruling out the values of 0.14 and 0.035 for this ratio (top and bottom panels of Figure~\ref{FCS_abund}) leads to an estimated  precision that is approximately 50\%. This estimate of the abundance ratio is significantly larger than the photospheric abundance ratio, $A({\rm Na})/A({\rm Ne}) = 0.02$ \citep{asp09}, and suggests an enhancement of about 3 to 4 in the abundance of Na in the corona if the active region plasma is coronal in origin.

For the Parkinson rocket spectra, the emission arises from a lower-temperature plasma indicative of a non-flaring active region. Figure~\ref{Parkinson_synth_sp} shows the KAP crystal scans compared with four theoretical spectra with $T_e = 1$, 1.5, 2, and 2.5~MK and $A({\rm Na})/A({\rm Ne}) = 0.07$. At $T_e \leqslant 2$~MK,  the ratio of the \ion{Na}{x} $w$--\ion{Ne}{ix} $w4$ blend at 11.002~\AA\ to the \ion{Na}{x} $z$ line at 11.191~\AA\  is sensitive to $T_e$, largely through the contribution of the \ion{Na}{ix} satellites $j$ and $k$ to \ion{Na}{x} line $z$. The ratio of the sum of these satellites to \ion{Na}{x} line $z$ over the temperature range 1~MK to 2.5~MK varies from 2.5 to 0.14. Of the four theoretical spectra shown, the one with $T_e = 1$~MK best fits the KAP spectrum. Although only just significant (total photon counts per 0.025~s equal to about 16), the feature at 11.15~\AA\ can be identified with the blend of \ion{Na}{ix} satellites $q$ and $r$; the appearance of these satellites confirms the temperature of $\sim 1$~MK. This conclusion is also supported by the gypsum crystal scan though this has lower statistical quality. In the original analysis of \citet{par75}, an emission measure distribution peaking at 3~MK was found, rather more than is found here with improved atomic data. The comparatively low temperature for this active region that we find is in keeping with the fact that solar activity was at a rather low level when the observations were taken.

If the temperature of the active region is only $\sim 1$~MK, the \ion{Fe}{xvii} lines at 11.023~\AA, 11.129~\AA, and 11.250~\AA\ should not be so strong. The ionization fractions of \citet{bry09} indicate that there is a negligible fraction of Fe$^{+16}$ ions for $T_e \leqslant 1.5$~MK (Figure~\ref{GofT_functions}). As with the FCS spectra, a higher-temperature component might be present. However, in that case the \ion{Na}{x} and \ion{Ne}{x} line emission would be much higher,  the fitted temperature would be correspondingly higher, and disagreements in the fit to the \ion{Na}{x} $w$--\ion{Ne}{ix} $w4$ blend and the weak 11.15~\AA\ line feature, formed by satellites $q$ and $r$, would result. There is a possibility, then, that the Fe$^{+16}$ ion fractions at $T_e \leqslant 1.5$~MK are in error, and that there is in fact a non-negligible fraction of Fe$^{+16}$ ions at very low temperatures.

\section{Conclusions}

In this work, recent collisional excitation results for \ion{Na}{x} X-ray line emission have been applied to solar spectra for the first time, together with new atomic data for \ion{Na}{ix} satellites, formed by both dielectronic recombination and inner-shell excitation. From these and other data from the {\sc chianti} atomic code, synthetic spectra were calculated as a function of electron temperature $T_e$. To date, the only observed spectra taken with high-resolution spectrometers in the 10.9--11.3~\AA\ range are from the {\it SMM} Flat Crystal Spectrometer (a single scan during the decay of a M1.5 flare) and two rocket-borne scanning crystal spectrometers in 1971 viewing a non-flaring active region. Comparison of the FCS and calculated spectra in the 10.9--11.3~\AA\ range shows that the \ion{Na}{x} X-ray lines with nearby \ion{Na}{ix} dielectronic satellites yield a Na/Ne abundance ratio through the blend of the \ion{Na}{x} $w$ (resonance) line with the \ion{Ne}{ix} $w4$ ($1s^2\,\,^1S_0 - 1s4p\,\,^1P_1$) line, these lines having almost identical contribution functions $G(T_e)$ for the low temperatures considered here. The value obtained, $A({\rm Na})/A({\rm Ne}) = 0.07 \pm 50$\%, is the coronal value of \citet{fel92a} to within uncertainties. It is significantly higher than the photospheric value, 0.020, and suggests that sodium (like other low-FIP elements) is enhanced by a factor of 3 to 4 in the corona. The lower-temperature observations obtained by \citet{par75} are best fitted with theoretical spectra having $T_e = 1$~MK, and are again consistent with a coronal Na/Ne abundance ratio. The presence of \ion{Fe}{xvii} lines in these spectra is unexpected, the most recent ionization equilibrium calculations of \citet{bry09} indicating either that there is a higher-temperature component present or (more likely) that there is a negligible fraction of Fe$^{+16}$ ions at $T_e \leqslant 1$~MK. A revision of ionization or recombination rates may be needed.

\begin{acknowledgements}
K. M. Aggarwal acknowledges financial support from EPSRC. The work of E. Landi is supported by several NASA grants. F. P. Keenan is grateful to AWE Aldermaston for the award of a William Penney Fellowship.  We are grateful to Dr Alexander Kramida for help and advice with running the adapted Cowan HFR program. Professor John Parkinson is also thanked for help and advice on his spectra, as is Dr M. F. Gu for advice on the Flexible Atomic Code used in this work.

\end{acknowledgements}

\end{document}